\documentclass[a4paper,10pt]{iopart}
\usepackage[utf8]{inputenc}
\usepackage{graphicx}
\usepackage{hyperref}
\usepackage{xcolor}

\newcommand{\SI}{\textrm{\tiny SI}}
\newcommand{\SRS}{\textrm{\tiny SRS}}
\newcommand{\C}{\mathcal{C}}

\newcommand{\Ni}{N_i}
\newcommand{\rhom}{\rho^\textrm{\tiny m}}
\newcommand{\nm}{r}
\newcommand{\rhof}{\bar\rho}
\newcommand{\nuf}{\bar\nu}

\begin{document}

\title{On a definition of the SI second with a set of optical clock transitions}
\author{Jérôme Lodewyck}
\address{LNE–SYRTE, Observatoire de Paris, Université PSL, CNRS, Sorbonne Université, 61 avenue de l'Observatoire, F-75014 Paris, France}
\ead{jerome.lodewyck@obspm.fr}

\begin{abstract}

The current SI second based on the atomic hyperfine transition in the ground state  of $^{133}$Cs is expected to be replaced by a new definition based on optical frequency standards, whose estimated uncertainty has now been established two orders of magnitude lower than the accuracy of the best Cs primary standards. However, such a redefinition of the second is hindered by the fact that many atomic species are potential contenders to become the new primary frequency standard.
In this paper, we propose to resolve this issue by defining a composite frequency unit based on the weighted geometric mean of the individual frequencies of different atomic transitions. This unit has the property to be realisable with any single clock whose transition composes the unit, provided that at least a few frequency ratios are available, with an accuracy that marginally differs from the nominal clock uncertainty. We show that the unit can be updated as the performances of the contributing transitions evolve, without incurring a drift on the unit itself.
\end{abstract}

\section{Introduction}

Since 1967, the definition of the Syst\`eme International (SI) unit of time, the second, has been based on the hyperfine energy splitting of the ground state of $^{133}$Cs atoms~\cite{8331917}. The initial definition, adopted at the 13th Conférence Générale des Poids et Mesures (CGPM) states that:
\begin{quote}
	The second is the duration of 9\,192\,631\,770 periods of the radiation corresponding to the transition between the two hyperfine levels of the ground state of the cesium 133 atom.
\end{quote}
Since then, the accuracy of best realisations of this definition, that is to say the uncertainty with which the unit is realised, has improved by one order of magnitude every decade, from the first cesium beam clocks to the modern cold atom fountain clocks. To take into account the evolution of these cesium standards, now reaching a control of systematic effects down to a few part in $10^{-16}$, the definition of the SI second was amended in 1997:
\begin{quote}
	This definition refers to a caesium atom at rest at a temperature of 0~K.
\end{quote}
The 26th CGPM in 2018 introduced a new wording of the definition of the time unit, in order to make it consistent with the newly adopted definitions of the other base units:
\begin{quote}
	The second, symbol s, is the SI unit of time. It is defined by taking the fixed numerical value of the caesium frequency $\Delta\nu_\textrm{\tiny Cs}$, the unperturbed ground-state hyperfine transition frequency of the caesium 133 atom, to be 9\,192\,631\,770 when expressed in the unit Hz, which is equal to s$^{-1}$.
\end{quote}
The SI is thus formally defined by fixing fundamental constants, among which $\Delta\nu_\textrm{\tiny Cs}$:
\begin{quote}
	The International System of Units, the SI, is the system of units in which the unperturbed ground state hyperfine transition frequency of the caesium 133 atom $\Delta\nu_\textrm{\tiny Cs}$ is 9\,192\,631\,770 Hz [\ldots]
\end{quote}
In this major update of the SI, the definition of the second is therefore left unchanged. However, since the late 2000s, clocks based on optical transitions in ions or neutral atoms have been developed with a control of systematic effects better than the accuracy of the best Cs standards~\cite{RevModPhys.87.637}, raising the question of a redefinition of the SI second~\cite{RIEHLE2015506, 8521703, 1742-6596-723-1-012053, margolis2014timekeepers}. A roadmap crafted by the Comité Consultatif Temps Fréquence (CCTF) of the Bureau International des Poids et Mesures (BIPM) and summarized in~\cite{0026-1394-55-2-188}, fixes milestones that the optical clocks must reach in order to make such a redefinition possible. Beyond the realisation of optical clocks with a low uncertainty, the roadmap emphasizes on the necessity to connect the new unit of time to the current SI second, to show the reproducibility of the optical clocks across different laboratories, and to measure frequency ratios between different optical transitions with an uncertainty matching the uncertainty of the best clocks. The progresses in completing these milestones let us envision a possible redefinition of the SI second as early as 2026. However, while the choice of cesium in 1967 was univocal\footnote{Retrospectively, the choice of cesium turned out to be the correct one as, in 2019, Cs is still one the most appropriate species for a low uncertainty microwave clock, equally matched with rubidium.}, current optical clocks are being developed with various atomic species, and various technologies. This diversity of high performance optical clocks is a wealth for the community, because it fosters applications of precise metrology in the field of fundamental physics. But it also means that none of these optical transitions currently stands out as an obvious choice for a new definition of the SI second, and the fast evolution of the field of optical frequency metrology makes picking a single atomic species uncertain. Facing this situation, it seems unavoidable that a new definition of the SI second should not designate a specific atomic transition as the new standard, but rather define the frequency unit from a weighted mix of the best realised optical transitions. This article proposes such a definition for a unit of time that can accommodate with the multiplicity of frequency standards and their evolution with time.

\section{Current status of the Cs atom in the SI}

\begin{figure}
	\begin{center}
		\includegraphics[width=7cm]{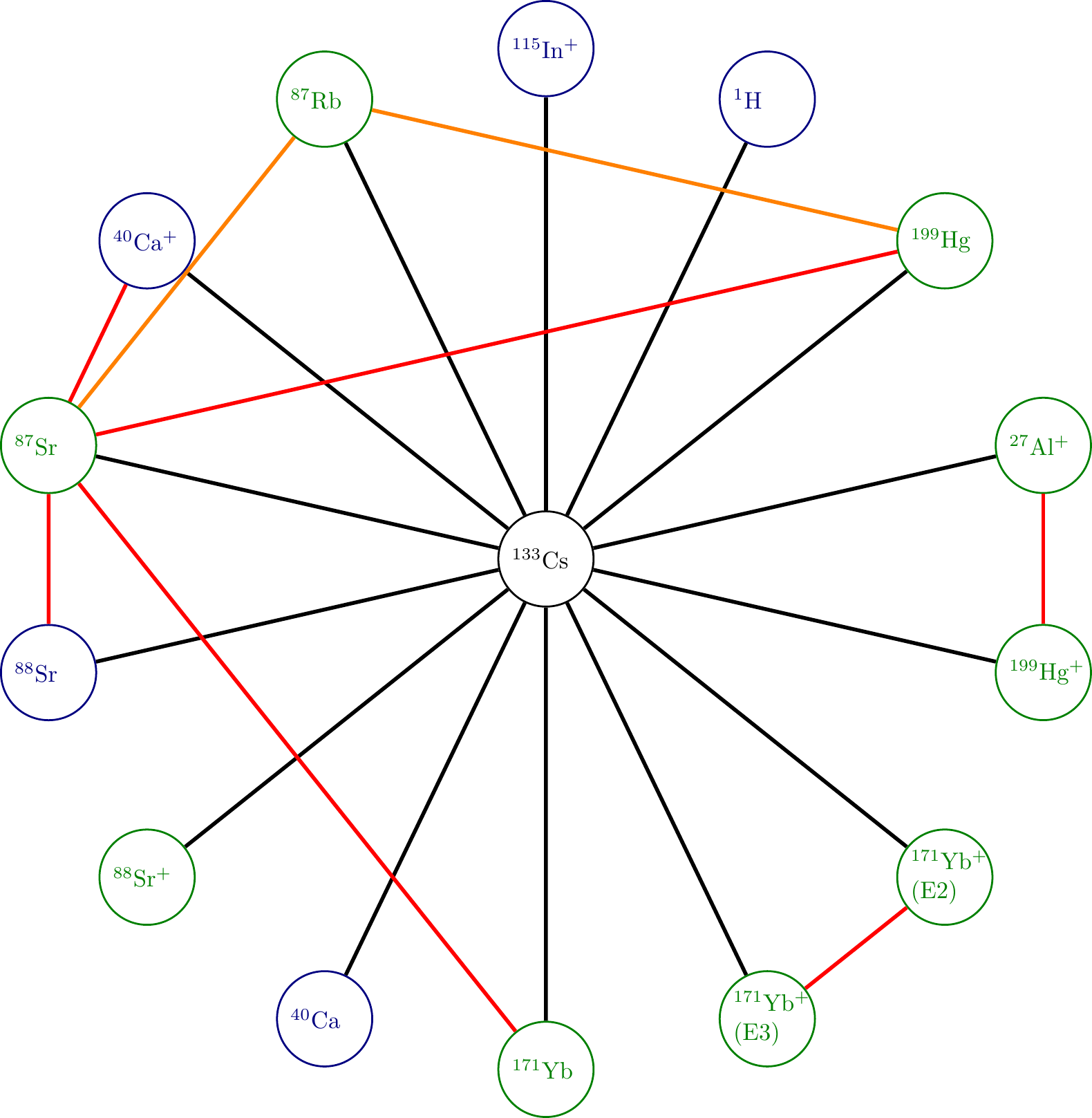}
		\caption{\label{fig:cslink}Measured frequency ratios, adapted from~\cite{0026-1394-55-2-188}. Green labels: secondary representations of the second (SRS). Blue: other optical transitions; Black lines: absolute frequency measurements against Cs; Red lines: optical-to-optical frequency ratios; Orange lines: optical-to-microwave frequency ratios against the Rb SRS.}
	\end{center}
\end{figure}

Cesium is still the unique primary frequency standard (PFS) with which the SI second can be realised. However, over the past decade, the frequency ratios between cesium clocks and various other microwave or optical clocks have been precisely measured, with uncertainties limited by the best cesium fountains in the low $10^{-16}$. From these published measurements, it becomes practically feasible to realise the SI second with a very low uncertainty with other atomic species. This possibility is enacted by the Comité International des Poids et Mesures (CIPM), via the publication of recommended frequencies for so-called secondary representations of the SI second (SRS). The current system of primary and secondary representations of the SI seconds is reproduced in table~\ref{tab:SRS}.

\begin{table}
	\begin{tabular}{rcl}
Frequency/Hz  & Fractional uncertainty  & Transition \\ \hline\hline
9\,192\,631\,770           & exact                 & $^{133}$Cs ground state hfs \\
6\,834\,682\,610.904\,3126   & $6   \times 10^{-16}$ & $^{87}$Rb ground state hfs \\
429\,228\,004\,229\,873.0  & $4   \times 10^{-16}$ & $^{87}$Sr 5s$^2$ $^1$S$_0$ -- 5s5p $^3$P$_0$ \\
444\,779\,044\,095\,486.5  & $1.5 \times 10^{-15}$ & $^{88}$Sr$^{+}$ 5s $^2$S$_{1/2}$ -- 4d $^2$D$_{5/2}$  \\
518\,295\,836\,590\,863.6  & $5   \times 10^{-16}$ & $^{171}$Yb 6s$^2$ $^1$S$_0$ -- 6s6p $^3$P$_0$  \\
642\,121\,496\,772\,645.0  & $6   \times 10^{-16}$ & $^{171}$Yb$^{+}$ $^2$S$_{1/2}$ -- $^2$F$_{7/2}$ \\
688\,358\,979\,309\,308.3  & $6   \times 10^{-16}$ & $^{171}$Yb$^{+}$ 6s $^2$S$_{1/2}$ -- 5d $^2$D$_{3/2}$ \\
1064\,721\,609\,899\,145.3 & $1.9 \times 10^{-15}$ & $^{199}$Hg$^{+}$ 5d$^{10}$6s $^2$S$_{1/2}$ -- 5d$^9$6s$^2$ $^2$D$_{5/2}$ \\
1121\,015\,393\,207\,857.3 & $1.9 \times 10^{-15}$ & $^{27}$Al$^{+}$ 3s$^2$ $^1$S$_0$ -- 3s3p $^3$P$_0$ \\
1128\,575\,290\,808\,154.4 & $5   \times 10^{-16}$ & $^{199}$Hg 6s$^2$ $^1$S$_0$ -- 6s6p $^3$P$_0$ \\
\hline
	\end{tabular}
	\caption{\label{tab:SRS}Table of the frequencies of the Cs primary and secondary representation of the SI second, adopted at the 2017 CIPM under the advise of the CCTF. Reproduced from~\cite{0026-1394-55-2-188}}
\end{table}

More recently, several frequency ratios that do not involve the cesium clock transition have also been measured, some of them with a relative uncertainty lower than the accuracy of the best cesium clocks (see figure~\ref{fig:cslink}). To incorporate these measurements in the recommended values of the SRS, an adjustment procedure using a least square algorithm or graph theory~\cite{0026-1394-52-5-628,0026-1394-53-6-1272} is now implemented by the CCTF. While apparently preserving the central role of cesium as the primary standard -- the recommended frequencies of the SRS are given in the SI unit Hz -- these procedures actually make a step towards a Cs-free, decentralized, definition of the frequency unit. For instance, computing the ratio between the recommended frequencies, in Hz, of the strontium and mercury optical clock transitions gives an estimate of the real value of this ratio with an uncertainty lower than the accuracy of the cesium fountain clocks. From this point of view, the outcome of the adjustment procedures is not a set of frequencies in Hz, but more fundamentally a set of best estimates for the ratios $\rho_{ij}$ between the frequencies $\nu_i$ of the connected atomic species:
\begin{equation}
	\label{eq:rho}
	\rho_{ij} = \frac{\nu_i}{\nu_j}\qquad \textrm{with} \quad \rho_{ij} = \rho_{ik}\rho_{kj}.
\end{equation}
The transitivity condition stated at the end of equation~(\ref{eq:rho}) is true, by definition, for the actual frequency ratios. However, a set of physical measurements of these ratios may fail to verify the transitivity relation because of systematic and statistical uncertainty in the measurements. To circumvent this issue, and to ensure that the best estimates for frequency ratios are transitive, the adjustment procedures of~\cite{0026-1394-52-5-628,0026-1394-53-6-1272} set the transitivity relations as fixed constraints. In this picture, the numbers $\nu_i^\SRS$ listed in the first column of table~\ref{tab:SRS} can be formally reformulated as the transitive rational frequency ratio matrix in the ($^{133}$Cs, $^{87}$Rb, $^{87}$Sr, $^{88}$Sr$^{+}$, $^{171}$Yb, $^{171}$Yb$^{+}$, $^{171}$Yb$^{+}$, $^{199}$Hg$^{+}$, $^{27}$Al$^{+}$, $^{199}$Hg) basis:
\begin{equation}
	\label{eq:rhomatrix}
	\rho^\SRS_{ij} = \frac{\nu_i^\SRS}{\nu_j^\SRS}, \quad
	\textrm{and the }^{133}\textrm{Cs ground state hfs is } 9\,192\,631\,770~\textrm{Hz}.
\end{equation}
This matrix representation explicitly highlights that the recommendations for the SRS are in essence independent of the Cs frequency, the latter being factorized by the instantiation of the definition of the SI second. The uncertainty matrix on the recommended frequency ratios would then read:
\begin{equation}
	\label{eq:deltarho}
	\delta\rho^\SRS_{ij} = \rho^\SRS_{ij}\sqrt{u_i^2 + u_j^2},
\end{equation}
where $u_i$ is the relative uncertainty of the realisation of the clock transition $i$. In this equation, we assumed that the uncertainty on the frequency ratios is limited by the systematic uncertainty of the clocks, and therefore is not improved by accumulating frequency ratio measurements or by exploiting the redundancy between the coefficients of the frequency ratio matrix\footnote{In practice, the uncertainty given by equation~(\ref{eq:deltarho}) may not be met. Exploiting statistical independence in the measurement of frequency ratios can lead to a lower uncertainty. On the other hand, if the frequency ratio $\rho_{ij}$ was not directly measured but deduced from the product $\rho_{ik}\rho_{kj}$, the uncertainty may be larger than the uncertainty given by equation~(\ref{eq:deltarho}). Such cases will be considered later in this paper.}. While equation~(\ref{eq:rhomatrix}) is equivalent to the list of frequencies of table~\ref{tab:SRS}, the uncertainty matrix of equation~(\ref{eq:deltarho}), although it is not currently published along with the adjustment of SRS frequencies chosen by the CCTF, contains more information than the second column of table~\ref{tab:SRS}, because it also incorporates the uncertainty of frequency ratios not including the Cs clock transitions.

Because it does not directly focus on a single atomic transition, the  matrix representation of the recommendations published by the CIPM, given by equations~(\ref{eq:rhomatrix}) and~(\ref{eq:deltarho}), can be used to construct a new definition of the SI second that can be realised not only with a clock based on a specific, arbitrarily chosen transition, but rather by any high performance optical frequency standard. For this, we just have to reformulate the anchor to the Cs stated in the second part of equation~(\ref{eq:rhomatrix}). This is the aim for the next sections.

\section{Notations}

We note $\nu_i$ the physical frequency of the clock transition $i$. This is a quantity with a physical dimension, and as such, it is not experimentally accessible without a prior definition of the frequency unit. On the other hand, the frequency ratios $\rho_{ij}$ between these frequencies, as defined by equation~(\ref{eq:rho}) can be experimentally measured by comparing clocks. We note $\rhom_{ij}$ the outcomes of such measurements. They generally do not form a complete frequency ratio matrix as pairs of clock transitions may have never been directly compared, nor do they satisfy the transitivity relation, as they typically deviate from the genuine frequency ratios by the uncertainty of the clocks. However, we shall assume that they are connected, \emph{i.e.} there are no subsets of the frequency ratio measurements that do not share at least one clock transition. This requirement is necessary in order to be able to compute, from the measurements, a most likely frequency ratio matrix that is both complete and transitive~\cite{0026-1394-52-5-628,0026-1394-53-6-1272}. We note $\rhof_{ij}$ such a frequency ratio matrix, of which the matrix $\rho^\SRS_{ij}$ is an example.

As stated above, $u_i$ is the uncertainty of the clocks based on the transition $i$. To ease the reading, we sometimes call $u_i$ the uncertainty of transition $i$.

\section{Requirements for a unit}

We consider a frequency unit constructed from the frequencies of various atomic clock transitions. Such a frequency unit $\nu$ is mathematically defined by a function $F$ of $n$ individual frequencies $\nu_i$:
\begin{equation}
	\nu = F(\nu_1,\ldots,\nu_n).
\end{equation}
For instance, the current SI frequency unit is defined by
\begin{equation}
	\label{eq:nuSI}
	\nu^\SI \equiv 1~\textrm{Hz} = \frac{\nu_\textrm{\tiny Cs}}{N^\SI} \quad \textrm{with}\ N^\SI = 9\,192\,631\,770.
\end{equation}
Such a definition must fulfill two \emph{sine qua non} requirements. The frequency unit \textbf{must} be:
\begin{itemize}
	\item \textbf{realisable} in the form of a physical signal, \emph{i.e.} there must exist a practical device, accessible to various laboratories, whose output is a realisation of the frequency unit with an inaccuracy as low as possible.
	\item \textbf{backward compatible} with the previous definitions of the unit, \emph{i.e.} the relative frequency difference between a new definition of the unit and the previous definition must be lower than the accuracy of the best realisation at the moment of the change of definition.
\end{itemize}
The current definition of the SI second based on Cs fulfills these requirements: It is realisable with a cold atom atomic fountain clock, and the constant $N^\SI$ was chosen so that the atomic second defined with Cs would match the previous definition of the SI second based on the ephemeris time~\cite{PhysRevLett.1.105}.

Beyond these requirements, additional properties can be desirable for a frequency unit. The frequency unit \textbf{should} be:
\begin{itemize}
	\item \textbf{optimal}: no other unit can be better realised than the chosen frequency unit.
	\item \textbf{universal}: a clock based on any transition $i$ should be sufficient to realise the frequency unit within the uncertainty of the clock.
	\item \textbf{representative}: the lower the uncertainty of a transition, the larger the weight of the frequency of this transition in the definition of the unit.
	\item \textbf{evolutive}: the definition evolves with the uncertainties of the various clock transitions.
\end{itemize}
With the advent of various optical clocks with uncertainties smaller than the accuracy of the cesium fountain clocks, the current definition of the SI second does not fulfill anymore these properties.

\section{Construction of a weighted frequency unit}

We now propose to construct a new frequency unit suitable for optical transitions. For this, we write a general form for the definition of the unit of frequency as the arithmetic mean of the frequencies of different atomic clock transitions belonging to the set $\C$:
\begin{equation}
	\nu = \sum_{i\in\C} a_i\nu_i.
\end{equation}
The question then arises of choosing the best values for the coefficients $a_i$, both to ensure that the unit can be realised with the lowest possible uncertainty, and to allow for these coefficients to evolve with time as the uncertainty of the various transitions composing the unit improves, without introducing a drift in the unit.
Let us first express that transitions should contribute with different weights $w_i$, based on their uncertainty. A transition with a lower uncertainty will have a larger weight than a transition with a higher uncertainty. Taking into account that each transition has its own nominal frequency, this yields the relation between the $a_i$ coefficients:
\begin{equation}
	\label{eq:weight}
	a_i\rho_{ij}/a_j = w_i/w_j
\end{equation}
for all pairs $(i,j)$, which is a relation symmetric with the transformation $i \leftrightarrow j$ given that $\rho_{ij} = 1/\rho_{ji}$. It is straightforward to check that the $a_i$ coefficients defined as:
\begin{equation}
	a_i = \frac{1}{N} w_i \prod_{k\in\C} \rho_{ki}^{w_k},
\end{equation}
where $N$ is a numerical constant, satisfy equation~(\ref{eq:weight}), provided the weights $w_i$ are normalised:
\begin{equation}
	\sum_{k\in\C} w_k = 1.
\end{equation}
Then, the frequency unit can be rewritten as:
\begin{equation}
	\nu = \sum_{i\in\C} a_i\nu_i = \sum_{i\in\C} \frac{1}{N} w_i \prod_{k\in\C} \rho_{ki}^{w_k} \nu_i = \frac{1}{N} \sum_{i\in\C} w_i \prod_{k\in\C} \nu_{k}^{w_k} = \frac{1}{N} \prod_{k\in\C} \nu_{k}^{w_k}
\end{equation}
The frequency unit is thus the weighted geometric mean of the frequencies of the clock transitions. This property fits with the fact that, in essence, the ratios between frequencies are quantities more representative than the frequencies themselves, which are conventional and unit-dependent. This fact is also used in the fitting procedure of reference~\cite{0026-1394-53-6-1272}, through the use of the logarithms of frequency ratios.

\section{Definition of the frequency unit}

Using the results of the previous section, we can now formally define a unit of frequency as the weighted geometric mean of the frequencies of a set $\C$ of clock transitions:
\begin{equation}
	\label{eq:geom}
	\nu = \frac{1}{N}\prod_{i\in\C} \nu_i^{w_i},
\end{equation}
where $\nu_i$ are the individual transition frequencies, $N$ is a numerical constant, and $w_i$ the normalized weight of the clock transition $i$. The normalization of the weights ensures that $\nu$ has the dimension of a frequency.
The current Cs based SI unit of frequency, as expressed by equation~(\ref{eq:nuSI}), is a special case of equation~(\ref{eq:geom}) with $N = N^\SI$ and $w_i = \delta_{i,\textrm{\tiny Cs}}$.

The frequency unit~(\ref{eq:geom}) is \textbf{realisable} with a single clock based on any transition $i$, and thus \textbf{universal}, through the use of the frequency ratio matrix:
\begin{equation}
	\label{eq:nurealisation}
	\left.\nu\right|_i =\frac{\nu_i}{\Ni}\qquad\textrm{with}\quad \Ni = N\prod_{k\in\C} \rhof_{ik}^{w_k},
\end{equation}
where $\rhof$ is a measurement-based estimate of $\rho$ satisfying the transitivity property, for instance the matrix $\rho^\SRS$ constructed from the recommended frequency of the SRS already published by the CIPM. $N_i$ is the factor linking the frequency of the clock transition $i$ to the frequency unit: it can be viewed as the recommended frequency for the clock transition $i$. The $N_k$ numbers satisfy the relation:
\begin{equation}
	\frac{N_i}{N_j} = \rhof_{ij}.
\end{equation}

Then, the relative uncertainty on the realisation of the frequency unit with a clock based on transition $i$ is given by
\begin{equation}
	\label{eq:deltanurealisation}
	\left.\frac{\delta \nu}{\nu}\right|_i = \sqrt{\left(\frac{\delta\Ni}{\Ni}\right)^2 + u_i^2}.
\end{equation}
This uncertainty is composed of the uncertainty $u_i$ of the clock used to realise the unit, but also of the uncertainty on the independently determined quantity $\Ni$. The frequency unit is thus \textbf{non-optimal} as soon as the definition incorporates a transition beside the best transition of the moment.

\section{Trade-off optimality vs. universality}

The quantity $\Ni$ required to realise the unit with a clock based on transition $i$ can be computed from a compilation of world-wide frequency ratio measurements. Its uncertainty $\delta \Ni$ thus depends on which ratios are available, on the uncertainty of the clocks that were used to measure them, but also on the correlations between these ratios, that can occur if systematic effects shift the frequency of different clocks in the same way, or if some ratios were measured with the same clocks, possibly at the same time. A best estimate of $\Ni$ and its uncertainty $\delta \Ni$ can be extracted from a least-squares fit that takes as input the set of measured frequency ratios and their correlations, as explained in~\cite{0026-1394-52-5-628}. The \ref{sec:leastsquares} of this paper explicitly details this calculation.

In this section, we consider the relative uncertainty $\delta \Ni/\Ni$ for a few ideal cases, in order to theoretically quantify the deviation from optimality of the unit proposed in the previous section. We show that an appropriate choice of weighs $w_i$ leads to a realisation of the unit with a limited sub-optimality, by both accurate and less accurate clocks, resulting in a universal unit without significantly compromising its optimality.

\subsection{$n$ clock transitions with identical uncertainties}

We first assume the unit is composed of $n$ clock transitions, whose frequencies are realised by clocks with the same relative uncertainty $u$, and for which all frequency ratios have been evaluated with a relative uncertainty $\sqrt{2}u$. The weights of the clock transitions in the unit should therefore be identical, \emph{i.e.} $w_i = 1/n$. In this case, the deviation of the unit from optimality, characterized by $\delta \Ni/\Ni$, is at most the clock uncertainty $u$ if the ratio measurements are fully correlated, or goes to zero like $1/\sqrt{n}$ for $n$ large if the frequency ratio measurements are uncorrelated (see \ref{sec:leastsquares}). In practice, a set of measured frequency ratios would make $\delta \Ni/\Ni$ lay between these two bounds, such that the relative uncertainty on the realisation of the unit for a clock based on any transition $i$ would satisfy:
\begin{equation}
	u < \left.\frac{\delta \nu}{\nu}\right|_i < \sqrt{2}u.
\end{equation}
It differs at most by a factor $\sqrt{2}$ from optimality, and approaches optimality when a large number of uncorrelated frequency ratio measurements are available.

\subsection{Partial frequency ratio matrix}
\label{sec:incomplete}

\begin{figure}
	\begin{center}
		\includegraphics[width = 0.6\textwidth]{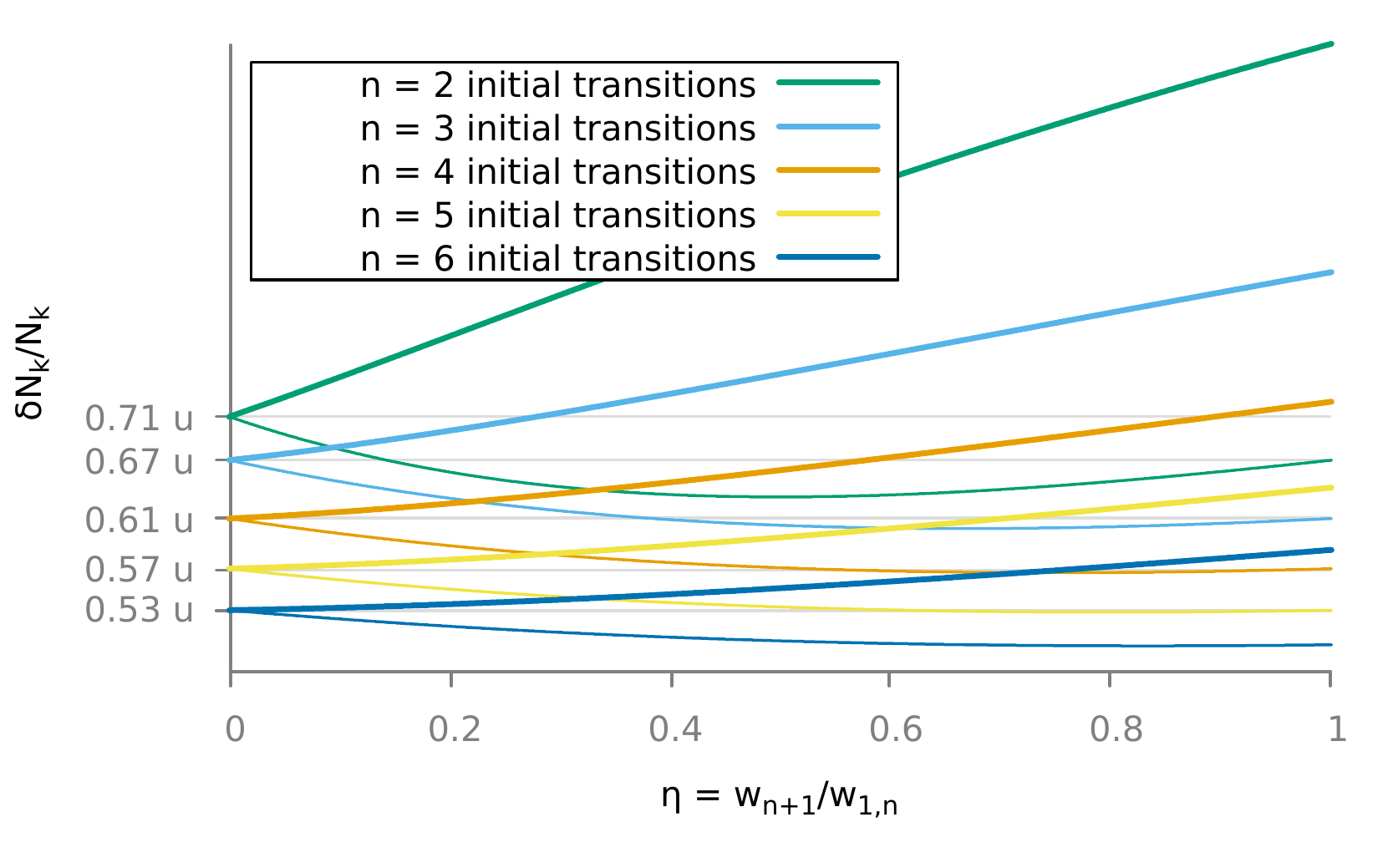}
		\caption{\label{fig:ncclocks} A transition $n+1$ is added to a pool of $n$ initial transitions. The frequency ratios between this new transition and the initial transitions is only available for a single initial transition, the ``connected transition''. This figure shows the uncertainty of the realisation of the frequency unit for the connected transition (thin lines) and unconnected transitions (thick lines), as a function of the weight $w_{n+1}$ of the newly introduced transition, from 0 (the unit does not incorporate the new transition) to $w_ {1,n}$ (the new transition has the same weight as the initial transitions). The measured frequency ratios are assumed to be uncorrelated.}
	\end{center}
\end{figure}

We now assume that another transition with the same uncertainty $u$ is added to the pool of transitions, bringing the total number of transitions to $n+1$. However, the connection of this new transition is not complete, \emph{i.e} there is no direct measurement of the frequency ratios between the additional transition and some of the initial $n$ transitions. To account for this, we may choose a different weight for the additional transition, say $w_{n+1} = \eta w_k$, $ 1 \leq k \leq n$. The addition of a partially connected transition to the initial pool of fully connected transition may result in an unwanted deviation from optimality in the realisation of the unit with the initial transitions, that we aim at quantifying.

The computation of the uncertainty $\delta \Ni/\Ni$ is reported in \ref{sec:leastsquares}. For completely correlated frequency ratio measurements, the addition of the new transition has the same effect as adding a fully connected transition, that is to say a marginal added uncertainty. For uncorrelated frequency ratio measurements, the uncertainty is plotted in Figure~\ref{fig:ncclocks} as a function of $\eta$ and $n$. It shows that for a pool of several transitions, the uncertainty of the realisation of the unit is not significantly altered by adding a partially connected transition to the unit, and that the degradation can be mitigated by setting a lower weight $\eta$ to the added transition. This allows to integrate the newly added transition into the unit, with an uncertainty on the realisation of the unit by this transition that is a decreasing function of $\eta$ and of the number of measured frequency ratios with the initial transitions.

\subsection{Two different clock transitions}

\begin{figure}
	\begin{center}
		\includegraphics[width = 0.6\textwidth]{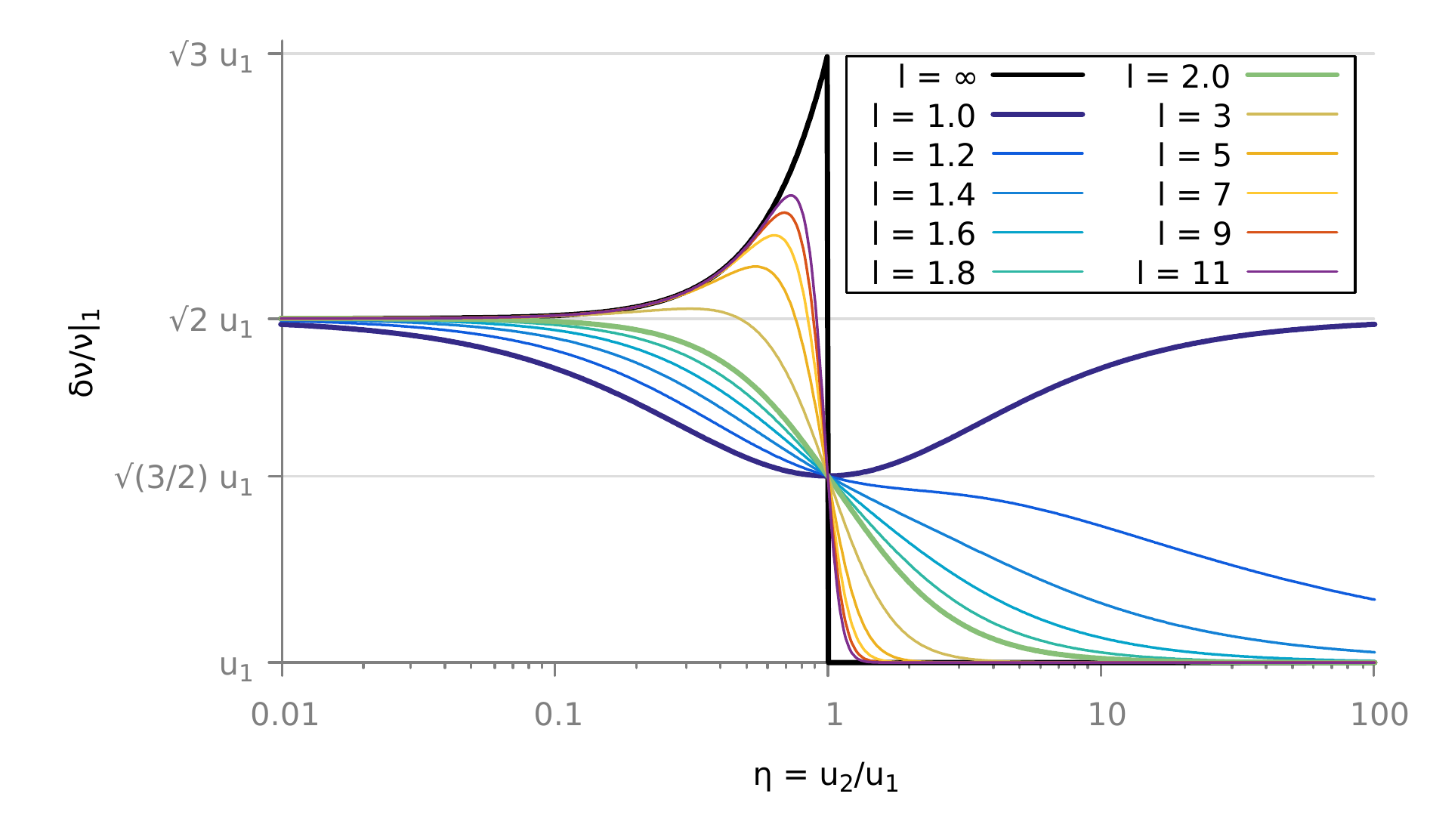}
		\caption{\label{fig:twoclocks}Frequency unit defined from a set of two transitions. The graph shows the relative uncertainty with which the unit can be realised with a clock based on transition 1, in units of $u_1$, as a function of $\eta = u_2/u_1$. The various curves represent different weighting functions $w_1 = \eta^lw_2$.}
	\end{center}
\end{figure}

We now consider the case in which the unit is composed of two clock transitions 1 and 2. We note $\eta = u_2/u_1$. The realisation of the frequency unit with a clock based on transition 1 reads:
\begin{equation}
	\label{eq:dnu2clocks}
	\left.\frac{\delta \nu}{\nu}\right|_1 = \sqrt{1 + w_2^2(1+\eta^2) }\ u_1.
\end{equation}
We now have to decide on the weight to attribute to each clock transition. A first choice would be to set $w_1 = \eta w_2$. However, this choice is such that the uncertainty on the realisation of the frequency unit with a clock based on transition 1 given by equation~(\ref{eq:dnu2clocks}) is minimal for $\eta = 1$, at $\sqrt{3/2}\,u_1$, and then increases to $\sqrt{2}\,u_1$ for $\eta$ going to infinity. This situation is not suitable, as adding a clock transition with a large uncertainty would degrade the ability to realise the frequency unit with a low uncertainty clock. To remedy this problem, one can choose $w_1 = \eta^l w_2$ with $l > 1$. In such a case, $\left.\delta \nu/\nu\right|_1$ goes to $u_1$ for $\eta \gg 1$ and to $\sqrt{2}\,u_1$ for $\eta \ll 1$. Now, the uncertainty of the realisation of the unit with the best clock is only limited by the uncertainty of this clock, while realising the unit with the worse clock is $\sqrt{2}$ the clock uncertainty (one uncertainty to for the connection to the unit \emph{i.e.} measuring the frequency ratio, one uncertainty for the realisation). These results are represented in figure~\ref{fig:twoclocks}.

\section{Evolution}

In the previous section, we showed that the weight of each transition in the definition of the unit should be adapted to the uncertainty of this transition. As the performances of the clocks improve, and as new transitions are evaluated, we may consider altering the weighting coefficients $w_i$. In doing so, the unit must be backward compatible. For this, an updated value of the normalisation constant $N$ must be calculated from the evolution of the weights and the transitive frequency ratio matrix $\rhof$ available from measurements.

We denote with a superscript $^{(m)}$ the value of the defining constants $N$ and $w_k$ at the $m$th evolution of the unit. Equating the realisation of the unit for an arbitrary transition $i$, given by equation~(\ref{eq:nurealisation}), before and after the change yields:
\begin{equation}
	\label{eq:evolN}
	N^{(m+1)} = N^{(m)}\ \frac{\prod_{k\in\C} \rhof_{ik}^{w_k^{(m)}}}{\prod_{k\in\C} \rhof_{ik}^{w_k^{(m+1)}}}.
\end{equation}
Using the transitivity property of the matrix $\rhof$, it is straightforward to check that $N^{(m+1)}$ does not depend on the choice of $i$.

Because the values of the measured frequency ratios $\rhof$ contain statistical noise, Equation~(\ref{eq:evolN}) can be interpreted as a random walk of the logarithm of $N$. As such, the unit may drift away from a previous definition more than the uncertainty of its realisations. However, according to Kolmogorov's three-series theorem, the random walk converges if the series
\begin{equation}
	\label{eq:kolmo}
	\sum_{m = 0}^\infty\ v_m \qquad \textrm{with} \quad v_m =  \sum_{k\in\C}\ \left[u_k^{(m+1)}\left(w_k^{(m+1)} - w_k^{(m)}\right)\right]^2
\end{equation}
converges over $m$. This condition is easily met if the unit is updated when the terms $v_m$ go to zero faster than $1/m$, and \emph{a fortiori} if they improve by a fixed factor between each update of the unit. The expression of $v_m$ in equation~(\ref{eq:kolmo}) is valid for a complete set of fully correlated frequency ratio measurements, but equation~(\ref{eq:vm}) can be used to compute $v_m$ for any given covariance matrix resulting from the fit of a set of frequency ratio measurements.

In order to illustrate the convergence of the unit, we conduct a numerical simulation of its evolution. For this, we fix exact frequency ratios. At each evolution step $m$, we draw random frequency ratios $\rhom$ that deviate from the exact ratios by the uncertainty set for each transition at step $m$. From these ratios, we estimate the most probable transitive frequency ratio matrix $\rhof$ with a least square method. We then use this ratio matrix to update the normalisation constant of the unit $N$ according to equation~(\ref{eq:evolN}). Equipped with the knowledge of the exact frequency ratios that our simulation provides us with, we then calculate the frequency of the unit and the accuracy of its realisation with clocks based on each transition using equations~(\ref{eq:geom}) and~(\ref{eq:nurealisation}) respectively.

\begin{figure}
    \begin{center}
		\includegraphics[width=\textwidth]{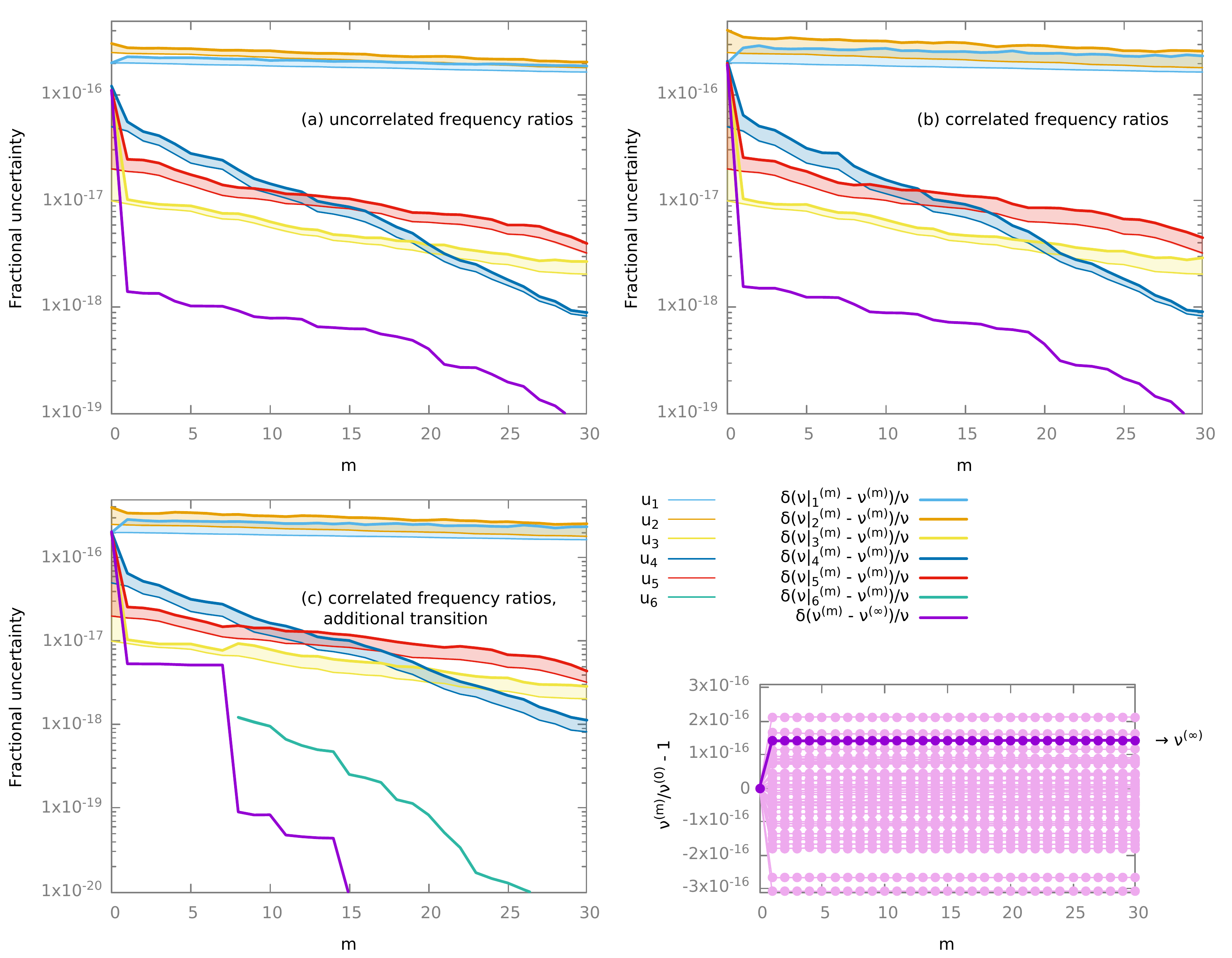}
		\caption{\label{fig:evol}Simulation of the evolution of the frequency unit, starting with an initial configuration ($m = 0$) in which a single transition with an uncertainty $u_1 = 2\times 10^{-16}$ contributes to the unit, \emph{i.e} reproducing the current SI second. At the first evolution step ($m = 1$), we start incorporating another microwave clock transition (labeled 2) and three optical clock transitions (labeled 3 to 5) to the unit. Each clock transition contributes to the unit with a weight $w_i \propto 1/u_i^2$. We model a possible stochastic evolution of the transitions' uncertainty with only marginal improvement for the microwave transitions, and different rates of improvement for the three optical transitions.\newline
		\textbf{Bottom right}: Example trajectories of the unit, with respect to its initial definition, corresponding to different measurements of frequency ratios. At $m = 1$, the unit experiences a step after the introduction of the new transitions. However, this step lies withing the uncertainty of the first transition initially defining the unit. After this first step, the unit converges fast to an asymptotic value, noted $\nu^{(\infty)}$ and represented for a specific highlighted trajectory.\newline
		The \textbf{mains graphs} show statistics over a large number of trajectories, for three different configurations in which all the frequency ratios between all transitions have been measured. For $a)$ the measurements are independent and limited by statistical noise; For $b)$ and $c)$ the measurements are maximally correlated (\emph{i.e.} limited by common systematic effects, or simultaneous); For $c)$, a sixth clock transition (labeled 6) with a radically improved uncertainty is introduced after step $m=7$. For all graphs, the thin lines show the transition uncertainties $u_i$ chosen for the simulation. The thick lines show the relative standard deviation in the realisation of the unit with a clock based on transition $i$, \emph{i.e} $\left.\delta \nu/\nu\right|_i$, matching its analytical expression given by equation~(\ref{eq:deltanurealisation}).
		The colored area between the thin and thick line thus represents the overhead due to the non-optimality of the unit.}
	\end{center}
\end{figure}
\begin{figure}
	\caption{
	\hspace{-1.8cm}\fboxsep=0.6ex \fboxrule=0mm \fcolorbox{black}{white}{\raisebox{1.2ex}{\makebox[1.57cm]{\null}}}\hspace{-1.8cm}
	(continued caption of Figure~\ref{fig:evol}) For the initial configuration $m = 0$, this area is null for the unique primary frequency standard (PFS, $i = 1$) defining the unit, and large for the other transitions acting as SRS. In case $b)$ and $c)$, even though the uncertainty of the SRS is better than the uncertainty of the PFS, the former can only  realise the unit with an uncertainty limited by the accuracy of the latter, as one can expect. However, in case $a)$ the statistical independence between the frequency ratio measurements enables a better realisation of the unit with SRS.\newline
	As soon as more transitions are introduced ($m > 0$), the thickness of the area is small compared to the respective transition uncertainty, meaning that it is possible to realise the unit with a clock based on any transition with an uncertainty very close to the nominal uncertainty of the clock. For $c)$, the newly added transition with a much better uncertainty becomes \emph{de facto} the new PFS, and the overhead in realising the unit with the other clock transitions increases.\newline
	The thick purple line shows the relative standard deviation of the unit with respect to its asymptotic value output from the simulation; it matches its analytical expression $\left(\sum_{m' = m}^\infty v_{m'}\right)^{1/2}$ where $v_{m'}$ is given by equation~(\ref{eq:vm}). In the correlated cases $b)$ and $c)$, $v_{m'}$ reduces to the expression of equation~(\ref{eq:kolmo}). From this equation, we expect that if the weights $w_i$ smoothly evolve, \emph{i.e.} $|w_k^{(m+1)} - w_k^{(m)}| \ll 1$, the change in the unit can be significantly lower than the uncertainties of the transitions $u_i$. This effectively appears in the simulations $a)$ and $b)$. In $c)$ however, the abrupt change in the weights due to the sudden introduction of the additional transition brings the change in the unit closer to the uncertainty of the transitions.}
\end{figure}

In a first simulation, we keep the transition uncertainties constant, but the weight of each transition is randomly redistributed at each update. This simulation shows the expected random walk behavior of the unit as the series~(\ref{eq:kolmo}) does not converge. In a second simulation, we show the convergence of the unit when the transition uncertainties are randomly improved at each update of the unit (Figure~\ref{fig:evol}).

\section{From a single primary frequency standard to multiple clock transitions}

Let us consider the special case for which $w_{i_0} = 1$ for a specific clock transition $i_0$, and $w_i = 0$ for all other transitions. In this case, the $\Ni$ factors reduce to:
\begin{equation}
	N_{i_0} = N \qquad \textrm{and} \qquad \Ni = N\rhof_{ii_0} \quad \textrm{for } i \neq i_0,
\end{equation}
such that the frequency unit can be directly realised with a clock based on the transition $i_0$ without additional uncertainty beside the uncertainty of the clock, while it can be realised with clocks based on other transitions with the added uncertainty on the frequency ratio $\bar \rho_{ii_0}$. This situation is precisely the current situation, in which we have a primary frequency standard (PFS) $i_0$ with which the frequency unit is defined by $\nu = \nu_{i_0}/N_{i_0} = \nu_{i_0}/N$ exactly, and secondary representations of the second (SRS) whose recommended frequency is $\nu_i^\SRS = N_i\nu = N\rhof_{ii_0} \nu$. In that view, the unit proposed in this paper is not an alternative to the current definition of the SI second with a single PFS and a set of SRS, but rather a generalisation thereof. This distinction has two important consequences.

First, if at some point in the future, a specific clock transition $i_0$ stands out with realisations significantly more accurate than the realisations of other transitions, the unit will evolve, according to the procedure described in the previous section, to the weights $w_i = \delta_{i,i_0}$. The unit would thus effectively fall back to a definition based on a single PFS, while the other transitions would acquire the status of the current SRS. However, this situation is reversible, should the other transitions catch up with the transition $i_0$.

Second, all the algorithms or procedures already in place designed to make use of data produced by primary and secondary standards can be readily adapted to the unit proposed here, by replacing the recommended frequency $\nu^\SRS=N^\SI\rhof_{i,\textrm{\tiny Cs}}$ and its uncertainty for SRS by $\Ni$ and its uncertainty for all transitions. A notable example is the steering of TAI (Temps Atomique International) by contributing atomic clocks. While TAI was historically steered by the $^{133}$Cs primary standards, contributions from $^{87}$Rb, and more recently from optical clocks (based on neutral $^{87}$Sr and $^{171}$Yb) are now incorporated in TAI, using the recommended frequency for these SRS, and adding its uncertainty to the clock uncertainty~\cite{wolf2006comparing,petit2015international}. With the new unit proposed in this paper, all contributing clocks would be steering as SRS are steering now. Namely, one would produce the frequency unit by scaling the frequency of a clock based on the atomic transition $i$ by the factor $1/\Ni$, as expressed by equation~(\ref{eq:nurealisation}), and use it to calibrate the frequency of the local oscillator used as a pivot to connect to TAI. The relative uncertainty on $\Ni$ would be used to fill the $u_{\tiny S\textrm{rep}}$ entry of the circular T. With the definition of the time unit proposed here, TAI would thus remain a truly atomic time, in the sense that all atomic transitions would contribute to TAI on an equal footing, with an added uncertainty $\delta N_i/N_i$ expressing how well the transition is connected to the unit.

\section{Practical realisation}

The practical definition of the unit requires a designated body in charge of deciding of the list of clock transitions contributing to the unit and their respective weights. The Frequency Standards Working Group (WGFS) of the CCL (Comit\'e Consultatif des Longueurs) - CCTF has defined a set of requirements and guidelines for the inclusion of transitions and frequency ratio measurements to obtain the list of recommended frequencies, the best of which being promoted to SRS~\cite{0026-1394-55-2-188}. It takes into account published frequency ratio measurements and sets an uncertainty for these ratios that depends on the published uncertainty, but also on the level of confidence provided by the repeatability of the measurements by different laboratories, and possible correlations between the frequency ratios. These conservative guidelines ensure that the set of SRS is not affected by possible outliers.

This very procedure can be used as is to set the weight of each transition that compose the unit proposed in this paper, and therefore to set the definition of the unit.
In addition to setting the definition itself, the working group could recommend values for the quantities $\Ni$ and their uncertainty $\delta \Ni$ required to realise the frequency unit with a clock based on transition $i$. The former can already be computed from the published values of the recommended frequencies for SRS, while the latter can be deduced from the covariance matrix of the adjusted frequency ratios, a by-product of the fitting procedure. Using these numerical constants, any laboratory equipped with a clock based on one of the transitions composing the unit would be able to realise it.

The shift to the unit proposed in this paper would thus comprise the following steps:
\begin{enumerate}
	\item The definition of the SI second is changed to:
	\begin{quote}
		The second, symbol s, is the SI unit of time. It is defined by taking the fixed numerical value of the weighted geometric mean of the frequencies of a set of atomic clocks transitions, to be a constant $N$ when expressed in the unit Hz, which is equal to s$^{-1}$. The weights and the constant $N$ are published by the CIPM and updated according to frequency ratio measurements, in order that the unit converges.
	\end{quote}
	\item At each meeting of the CCTF, a transitive list of recommended frequency ratios and their covariance matrix is published, along with the associated $\Ni$ factors and their uncertainties. These quantities are derived from a fit of all published frequency ratio measurements~\cite{0026-1394-52-5-628,0026-1394-53-6-1272}. As it is currently done for the recommended frequencies of SRS, additional safeguards can be applied in order to prevent outliers, such a requiring that frequency ratios should be directly or indirectly measured by at least two independent sets of clocks, or adding a margin to the published uncertainties.
	\item At each meeting of the CCTF, the possibility to update the weights of the clock transitions is evaluated. It could for instance be decided each time the quantity $v_m$ is reduced by a constant factor, such as two. This results in an exponential decay of the $v_m$ terms with $m$, thus ensuring the rapid convergence of their series. Such an update would notably occur at the first meeting, so that the weights and the normalisation constant would depart from their current value $w_i = \delta_{i,\textrm{\tiny Cs}}$ and $N = 9\,192\,631\,770$.
\end{enumerate}

The unit proposed in this paper is particularly relevant when a large number of frequency ratios between optical clock transitions have been measured, hence bringing it closer to optimality. The availability of such frequency ratios is now becoming a reality with optical clocks being compared through optical fiber links, such that changing the definition of the unit could be foreseen in the next few years. Yet, in order to illustrate the evolution process of the unit, the \ref{sec:annexsimul} shows how the unit would be now defined if it had been adopted before the 2015 meeting of the CCTF at which a few optical-to-optical frequency ratios were considered for the determination of the recommended frequencies of SRS.

The evolution step (iii) makes that the effective definition of the unit is regularly updated, by adopting new weights. One could thus argue that, through this process, the unit is constantly changing, making it indecisive. However, in essence, the unit is not changing: it keeps being the value $v^{(\infty)}$ whose existence is ensured by the convergence criterion that must be met, and each evolution step brings us closer to begin able to realise it, as illustrated by example trajectories of the unit shown on figure~\ref{fig:evol}. Admittedly, $v^{(\infty)}$ is accidental because it depends on the outcome of the frequency ratio measurements that drive the evolution of the unit. However, it is the case for all base units of the SI: the chosen values of fundamental constants are experimental accidents, but become defining once fixed.

\section{Are we fixing frequency ratios?}

The new SI adopted by the 26th CGPM emphasizes on the essence of fixing physical constants to define base units. For instance, the meter is defined by fixing the speed of light, the kilogram by fixing the Planck constant,\ldots. In this new SI, the wording of the definition of the SI second has been altered in order to make explicit that the unit of time is defined by fixing the frequency of the Cs ground state hfs to $N^\SI$, as reported in the introduction. Therefore, the question arises of which physical constant is fixed in the frequency unit proposed in this paper. One would be tempted to answer that frequency ratios are fixed because agreed-upon values of these ratios are necessary in order to realise the unit using equation~(\ref{eq:nurealisation}). Such a statement would be contradictory, because these frequency ratios are dimension-less, measurable quantities imposed to us by Nature, that we are thus not allowed to arbitrarily fix.

Indeed, it is crucial to notice that the \emph{definition} itself, given by equation~(\ref{eq:geom}) does not rely of the availability nor on the value of these ratios, but only on the arbitrarily fixable constants $N$ and $w_i$. Only the \emph{realisation} of the unit requires the knowledge of the frequency ratios. But it is already the case that measurable physical quantities are required in order to realise the SI second. For instance, the polarisability of the Cs atom is required to correct for the black-body radiation shift, the quadratic Zeeman shift coefficient is required to cancel this effect, the interaction strength between Cs atoms is required to compensate for collisional effects, \ldots Without prior measurements of these measurable physical quantities, it would not be possible to realise the SI second with Cs clocks. But the value of the Cs polarisability itself, although required by the definition,
is not stated nor mentioned in the definition of the SI second, which is above such mundanes. In the same spirit, the prior knowledge of measurable frequency ratios is required to realise the unit proposed in this paper, but they are not part of the definition, and \emph{a fortiori} not fixed by the definition. However, the availability, the uncertainty or the reproducibility of frequency ratios connecting a specific atomic transition to the pool of transitions already used to define the unit may influence the decision to incorporate or not this atomic transition to the pool. This requirement is aligned with the roadmap for the redefinition of the SI second~\cite{0026-1394-55-2-188}, which already requires that a large number of frequency ratios are measured and reproduced with a low uncertainty before a redefinition of the SI second can be agreed upon.

Unlike the other base units of the SI that are defined by directly fixing fundamental constants such as $c$, $h$, $e$, $k_B$, $N_A$, the Cs-based definition of the SI second only indirectly fixes a fundamental constant. Indeed, one can write the frequency of the Cs clock transition as the product of a constant with the dimension of a frequency, \emph{e.g.} the Rydberg constant $cR_\infty$, and a dimension-less function $\mathcal F_\textrm{\tiny Cs}$ of dimension-less physical constants that characterize the scale of the various physical interactions at stake in the Cs atom, such as the fine structure constant $\alpha$, the electron to proton mass ratio $m_e/m_p$, \ldots:
\begin{equation}
	\nu^\SI = \frac{\nu_\textrm{\tiny Cs}}{N^\SI} = \frac{1}{N^\SI}cR_\infty\,\mathcal{F}_\textrm{\tiny Cs}(\alpha, m_e/m_p,\ldots).
\end{equation}
Fixing the frequency of the Cs clock transition therefore amounts to fixing $R_\infty$ to the value $\nu^\SI N^\SI/c\mathcal{F}_\textrm{\tiny Cs}$, which is unknown because $\mathcal{F}_\textrm{\tiny Cs}$ is not exactly calculable, and its constant parameters only known with a finite uncertainty. While it would be conceptually more satisfying to fix a fundamental constant such as $cR_\infty$ to define the SI second, the choice of fixing the frequency of the Cs clock transition is driven by the practical availability of clocks able to realise the unit with the best accuracy.

This situation remains unchanged with the unit proposed in this paper, because, even though it fixes several constants ($N$ and $w_i$) instead of a single one ($\Delta\nu_\textrm{\tiny Cs}$), the frequency unit can still be expressed in the form:
\begin{equation}
	\nu = \frac{1}{N} \prod_{i\in\mathcal{C}}\nu_i^{w_i} = \frac{1}{N} cR_\infty\,\mathcal{F}(\alpha,m_e/m_p,\ldots).
\end{equation}
The function $\mathcal{F}$ simply becomes a combination of the different functions $\mathcal{F}_i$:
\begin{equation}
	\mathcal{F} = \prod_{i\in\mathcal{C}}\mathcal{F}_i^{w_i}.
\end{equation}
and thus remains unknown as well. However, this expression offers new possibilities. If the fundamental constants such as $\alpha$ are eventually found to drift with time~\cite{flambaum2009search}, one could envision to constraint the choice of the weights $w_i$ such that $\mathcal{F}$ is insensitive to variations of the fundamental constants. For instance, the Yb$^+$ E3 transition has a large negative dependence on $\alpha$ variations~\cite{flambaum2009search} that can be compensated for by the positive dependence of Sr, Yb, Al$^+$, Hg, at the expense of a reduced weight for the Yb$^+$ E3 transition.

\section{Summary and conclusion}

In summary, we showed that a frequency unit $\nu$ can be defined as the weighted geometric mean of the frequencies of a set $\C$ of clock transitions:
\begin{equation}
	\nu = \frac{1}{N} \prod_{i\in\C} \nu_i^{w_i}.
\end{equation}
The normalisation constant $N$ and the weights $w_i$ are defining constants for the unit. The choice of the latter is driven by the relative uncertainty $u_i$ of the transition with $w_i \propto 1/u_i^2$. The former is set in order to ensure the continuity with a previous definition of the unit. Such a unit is realisable with a single clock using a frequency ratio matrix satisfying the transitivity property~(\ref{eq:rho}), obtained from an adjustment of a connex but possibly incomplete set of experimentally determined frequency ratios between the transitions. The accuracy of the realisation is smaller than $\sqrt{2}u$ where $u$ is the uncertainty of the clock used for the realisation, and goes to $u$ for the best clocks of the pool when the size of the pool grows.

The backward compatible evolution of the definition of the second makes that technical evolution in the frequency standards do not change the essence of the definition of the frequency unit, but only the defining constants, in a way that only improves the uncertainty of the realisation, without introducing a drift of the unit.

Laboratories already equipped with high performance optical clocks will not be incited to reorient their research activities to implement a clock based on an arbitrarily chosen definition based on a single atomic species, thus avoiding that this atomic species becomes over-represented in metrology laboratories. On the contrary, defining the SI second with the unit proposed here would encourage laboratories to develop pairs of clocks never demonstrated before in order to improve the optimality of the unit. By doing so, the optical frequency community would preserve the diversity of atomic transitions currently being investigated, their enhanced potential, and their numerous applications in tests of fundamental physics.

Beyond this community, the SI is a field of physics apprehensible to the general public. It is therefore important that the choice of unit can be easily explained with a simple wording. This necessity became recently manifest, with the introduction of the new SI for which it is challenging to intuitively explain the relation between the Planck constant and the kilogram. Likewise, some aspects of the unit proposed in this paper rely on more abstract mathematical reasoning than the current definition solely based on the physical probing of Cs atoms. However, the complexity mainly lies in the proofs of correctness of the unit (optimality and convergence). The basic principle, on the other hand, is more accessible: in order to avoid an arbitrary and ephemeral choice among many possible candidates, the second is defined as the average frequency of the best realised atomic clocks transitions at any time, with evidence that the unit remains consistent over time. Such an explanation is arguably understandable while remaining correct.

\section*{Acknowledgments}

The author thanks S\'ebastien Bize for helpful suggestions about the manuscript.

\section*{References}

\bibliographystyle{unsrt}
\bibliography{SIsecond}

\clearpage
\appendix

\section{Least square adjustment of frequency ratios}
\label{sec:leastsquares}

We consider a set $\left\{\rhom_k,\ 1 \leq k \leq \nm\right\}$ of measured frequency ratios between the frequencies of different clock transitions. Because of statistical and systematic uncertainty in the measurements, these ratios deviate from the actual frequency ratios $\rho$, and do not fulfill the transitivity property~(\ref{eq:rho}). It is possible to derive a most likely transitive frequency ratio matrix $ \rhof$ from these measurements using a least squares procedure, as detailed in~\cite{0026-1394-52-5-628}. For this, one has to minimize the squared sum of the residuals:
\begin{equation}
	\label{eq:chi2}
	\chi^2 = \Delta^T W^{-1} \Delta,
\end{equation}
where $\Delta$ is the vector of size $\nm$ gathering the differences $\rhom_k - \rhof_k$ between the measured and most likely ratios, and $W$ is the $\nm\times\nm$ covariance matrix of the measured ratios. Because of the transitivity relations linking the coefficients of $\rhof$, the fitting procedure only involves $n-1$ free parameters, $n$ being the number of different clock transitions under consideration. A convenient choice is to arbitrarily set the frequency of a given clock transition $i_0$ to $\nuf_{i_0} = 1$, and to express the frequency ratios $\rhof_k$ as a function of the frequencies $\nuf_i$ of the $n-1$ other transitions.

This procedure yields the most likely values for the $\nuf_j$, either with a linear least squares algorithm (using a linearisation of the frequencies $\nuf_j$ around an \emph{a priori} guess)~\cite{0026-1394-52-5-628} or with a non-linear least squares algorithm (using the Jacobian of the system $J_{kl} = \partial \Delta_k/\partial \nuf_l$). The outcome is the best estimate for the frequency ratio matrix $\rhof_{ij} = \nuf_i/\nuf_j$, as well as the covariance matrix:
\begin{equation}
	\Sigma_{ij}= 2\left(\frac{\partial^2\chi^2}{\partial \nuf_i\partial \nuf_j}\right)^{-1}.
\end{equation}
This is a square matrix with size $n-1$, but in order to simplify the notations hereafter, we consider additional row and column at position $i_0$ with 0 components, giving a row and a column per clock transition. From this covariance matrix, on can deduce the uncertainty on the frequency ratios:
\begin{equation}
	\label{eq:deltarhoijml}
	\frac{\delta\rhof_{ij}}{\rhof_{ij}} = \sqrt{\frac{\Sigma_{ii}}{{\nuf_i}^2} + \frac{\Sigma_{jj}}{{\nuf_j}^2} - 2\frac{\Sigma_{ij}}{\nuf_i\nuf_j}}.
\end{equation}
To calculate the uncertainty on the geometric mean $\Ni = N \prod_{k} \rhof_{ik}^{w_k}$, it is convenient to choose the fixed frequency $i_0 = i$. In this case:
\begin{equation}
	\label{eq:deltaNi}
	\frac{\delta\Ni}{\Ni} = \sqrt{\sum_{j,l = 1}^n w_j w_l \frac{\Sigma_{jl}}{\nuf_j\nuf_l}} \qquad \textrm{for } i = i_0.
\end{equation}
Similarly,
\begin{equation}
	\label{eq:vm}
	v_m = \sum_{j,l = 1}^n \left(w^{(m+1)}_j - w^{(m)}_j\right) \left(w^{(m+1)}_l - w^{(m)}_l\right) \frac{\Sigma_{jl}}{\nuf_j\nuf_l}.
\end{equation}
Although some steps in the calculation of $\delta\rhof_{ij}/\rhof_{ij}$ and $\delta\Ni/\Ni$ depend on $i_0$, these quantities themselves are eventually independent of $i_0$.

We now make explicit these uncertainties in a few specific cases

\subsection{Uncorrelated complete set of measured ratios}

We first assume that the measured frequency ratios form a complete matrix (\emph{i.e.} all possible frequency ratios between the $n$ atomic species have been measured), and that all measurements are uncorrelated (\emph{i.e.} the measurement are independent, and limited by the statistical uncertainty or by systematic effects that are not in common mode between different measurements). In this very favourable case, there is an averaging effect between the measurements that produce a most-likely frequency ratio matrix $\rhof$ with an uncertainty lower than the individual uncertainty of the clocks. Explicitly, the input covariance matrix is diagonal and reads:
\begin{equation}
	W_{kk} = \delta\rho_{ij}^2 =  \rho_{ij}^2\left(u_i^2 + u_j^2\right),
\end{equation}
where $i$ and $j$ are the indexes of the two transitions involved in the frequency ratio measurement $k$, and $u_l$ is the relative uncertainty of transition $l$. Using the formulas above, one can then express the relative uncertainty on the most likely frequency ratios as a function of the individual transition frequencies. For instance, with a set $\mathcal{C}$ of $n = 3$ clock transitions, this uncertainty reads:
\begin{equation}
	\delta\rhof_{ij} = \sqrt{1 - \frac{u_i^2 + u_j^2}{2\sum_{k\in\mathcal{C} }u^2_k}}\ \delta\rho_{ij} \qquad \textrm{for } n = 3.
\end{equation}
This equation shows that the uncertainty $\delta\rhof_{ij}$ on the most likely frequency ratio $\rhof_{ij}$ is smaller than the uncertainty $\delta\rho_{ij}$ on the measured frequency ratio, by exploiting the redundant information in the uncorrelated frequency ratio matrix. For a set of $n$ transitions with the same uncertainty $u$, we have:
\begin{equation}
	\frac{\delta\rhof_{ij}}{\rhof_{ij}} = \frac{2}{\sqrt{n}}u,
\end{equation}
and
\begin{equation}
	\frac{\delta\Ni}{\Ni} = \frac{\sqrt{2(n-1)}}{n}u.
\end{equation}
The $1/\sqrt{n}$ behaviour of these uncertainties also illustrates the averaging effect.

\subsection{Fully correlated complete set of measured ratios}

We now make the opposite assumption: all possible frequency ratios between the various clock transitions are measured, but all these measurement are performed with the same set of clocks, at the same time. This means that statistic and systematic uncertainties are fully correlated. The diagonal terms of the covariance matrix $W$ are unchanged, but it now features off-diagonal terms.

For $n$ transitions with the same uncertainty $u$, this yields
\begin{equation}
	\frac{\delta\rhof_{ij}}{\rhof_{ij}} = \sqrt{2}u,
\end{equation}
and
\begin{equation}
	\frac{\delta\Ni}{\Ni} = \sqrt{1 - \frac 1 n}u.
\end{equation}
The uncertainty on all frequency ratios is limited by the transition uncertainties, and the relative uncertainty on $\Ni$ is at most the transition uncertainty $u$.

\subsection{Incomplete set of measured ratios}

We now detail the situation considered in section~\ref{sec:incomplete}. The complete frequency ratios are measured for a set of $n$ transitions, and a transition $n+1$, with the same uncertainty $u$ as the initial transitions is added to the pool of transitions, but only a limited number $n_c < n$ of frequency ratios involving this last transition have been measured. We define $\eta = w_{n+1}/w_{1 .. n}$  the relative weight of the additional transition in the definition of the frequency unit. With these assumptions, the additional relative uncertainty of the realisation of the frequency unit reads:

For uncorrelated measured frequency ratios:
	\begin{equation}
		\frac{\delta\Ni}{\Ni} = \frac{\sqrt{2}}{n + \eta}\sqrt{\frac{(n + n_c)\eta^2 + n(n_c-1)(2\eta - 1)}{n_c(n+1)} + n-1} \qquad \textrm{for } 1 \leq i \leq n_c
	\end{equation}
	\begin{equation}
		\frac{\delta\Ni}{\Ni} = \frac{\sqrt{2}}{n + \eta}\sqrt{\frac{n + n_c + 1}{n\,n_c}\eta^2 + 2\eta + n-1} \qquad \textrm{for } n_c < i \leq n
	\end{equation}
	\begin{equation}
		\frac{\delta N_{n+1}}{N_{n+1}} = \frac{\sqrt{2}}{n+\eta}\sqrt{\frac{n^2 + n - n_c}{n_c}}
	\end{equation}

For correlated measured frequency ratios:
	\begin{equation}
		\frac{\delta\Ni}{\Ni} = \frac{\sqrt{2\eta^2 + 2(n-1)\eta + n(n-1)}}{n + \eta} \qquad \textrm{for } 1 \leq i \leq n
	\end{equation}
	\begin{equation}
		\frac{\delta N_{n+1}}{N_{n+1}} = \frac{\sqrt{n(n+1)}}{n+\eta}
	\end{equation}

\section{defining the unit with the current frequency ratio measurements}
\label{sec:annexsimul}

\begin{table}
\footnotesize
\hspace{-1cm}
\begin{tabular}{@{}llll}
\bf 2015\\
\br
Species & $w_i$ & $N_i$ & $\delta N_i/N_i$\\
\mr
  $^{133}$Cs         & 0.089 & 9192631770          & $1.7\times 10^{-16}$ \\
  $^{171}$Yb         & 0.005 & 518295836590865.105 & $9.4\times 10^{-16}$ \\
  $^{171}$Yb$^+$(E2) & 0.071 & 642121496772645.04  & $2.7\times 10^{-16}$ \\
  $^{171}$Yb$^+$(E3) & 0.063 & 688358979309308.379 & $8.6\times 10^{-17}$ \\
  $^{199}$Hg         & 0.362 & 1128575290808154.79 & $8.6\times 10^{-16}$ \\
  $^{199}$Hg$^+$     & 0.010 & 1064721609899145.3  & $6.6\times 10^{-16}$ \\
  $^{27}$Al$^+$      & 0.011 & 1121015393207857.31 & $6.4\times 10^{-16}$ \\
  $^{87}$Rb          & 0.020 & 6834682610.904307   & $4.7\times 10^{-16}$ \\
  $^{87}$Sr          & 0.368 & 429228004229873.166 & $7.4\times 10^{-17}$ \\
\br
\end{tabular}
\hspace{0.6cm}
\begin{tabular}{@{}llll}
\bf 2017\\
\br
Species & $w_i$ & $N_i$ & $\delta N_i/N_i$\\
\mr
  $^{133}$Cs         & 0.081 & 9192631770.00000213 & $9.5\times 10^{-17}$ \\
  $^{171}$Yb         & 0.327 & 518295836590863.762 & $3.8\times 10^{-17}$ \\
  $^{171}$Yb$^+$(E2) & 0.016 & 642121496772645.189 & $2.6\times 10^{-16}$ \\
  $^{171}$Yb$^+$(E3) & 0.015 & 688358979309308.539 & $2.9\times 10^{-16}$ \\
  $^{199}$Hg         & 0.166 & 1128575290808154.68 & $6.4\times 10^{-17}$ \\
  $^{199}$Hg$^+$     & 0.002 & 1064721609899145.55 & $6.5\times 10^{-16}$ \\
  $^{27}$Al$^+$      & 0.003 & 1121015393207857.57 & $6.3\times 10^{-16}$ \\
  $^{87}$Rb          & 0.037 & 6834682610.90431418 & $1.6\times 10^{-16}$ \\
  $^{87}$Sr          & 0.353 & 429228004229873.134 & $2.7\times 10^{-17}$ \\
\br
\end{tabular}\\
\caption{\label{tab:simul}Proposition for the weights of the various transitions composing the unit, using the list of frequency ratio measurements published before 2015 (left) and 2017 (right), when the 20th and 21st meetings of the CCTF were held. We calculate $\sqrt{v_{\tiny 2015}} = 1.7 \times 10^{-16}$ at the first update of the unit, and $\sqrt{v_{\tiny 2017}} = 3.5 \times 10^{-17}$. The significant reduction of $v$ in 2017 with respect to 2015 justifies a new update of the unit, enabled by the inclusion of new optical frequency ratio measurements. The normalisation constant is set to $N = 203102222210226.561$ in 2015, and to $N = 152329318467266.642$ in 2017, in order to ensure continuity with the previous Cs based definition of the unit.}
\end{table}
\normalsize

The BIPM website publishes the list of measured frequency ratios that is currently being used as the input of the least-squares algorithm~\cite{0026-1394-52-5-628} used to determine the recommended frequency of the SRS. This list can be used as a playground to simulate the unit proposed in this paper, as if it had already been adopted. From the list of frequency ratios published before a given date, we determine the transitive set of most probable frequency ratios $\rhof_{ij}$ and their covariance matrix $\Sigma_{ij}$.

In order to define the unit one has to choose the weights $w_i$. In this paper, we prescribed to chose weights proportional to $1/u_i^2$ where $u_i$ is the uncertainty of the clocks realising the transition $i$. However, in practice, frequency ratios can be measured by different clocks using the transition $i$ with varying statistical and systematic uncertainties, possibly larger than the best uncertainty reported for transition $i$. $u_i$ is therefore ill-defined. Instead, the relevant uncertainty can be found in the values of $\delta \rhof_{ij}$ as given by the fit procedure, which represents how the clock transition $i$ is connected to other transitions by actual frequency ratio measurements. Here, we heuristically choose $w_i$ to be proportional to the average of the two largest values of $1/\delta \rhof_{ij}^2$ over $j$. This makes $w_i$ representative of the uncertainty of the best frequency ratio measurements with transition $i$, while avoiding that a single frequency ratio would exclusively steer the unit. One could alternatively choose the weights of the clocks with a specific optimisation procedure in order to \emph{e.g.} minimize the sub-optimality of the best transitions. All in all, these possible choices become similar as more frequency ratios between optical transitions are measured with state-of-the-art clocks; and any of these choices is definitely better than setting the full weight of the unit on a single atomic transition in the current and forseable context where many different optical clocks are developed.

From the weights $w_i$ and the covariance matrix $\Sigma_{ij}$, one can calculate the factors $N_i$ and their uncertainties, as well as $v_m$ and decide whether the unit should be updated or not. If so, the new normalisation constant $N$ is calculated. Table~\ref{tab:simul} reports on the outcome of such a procedure using the frequency ratio measurements published prior to 2015 and 2017.

\end{document}